\documentclass[prb,aps,showpacs,twocolumn,floatfix]{revtex4}

\usepackage{graphicx,color}
\usepackage{dcolumn}
\usepackage{bm}

\newcommand{\tonda}[1]{ \left( #1 \right) }
\newcommand{\graf}[1]{ \left\{ #1 \right\} }

\newcommand{\gr} [1]{\textbf #1}
\newcommand{\quadr}[1]{ \left[ #1 \right] }

\begin{document}

\title{\bf{Competitive density waves in quasi-one-dimensional electron systems }}
\author{
Samuele Bissola and Alberto Parola
}
\affiliation {
Dipartimento di Fisica e Matematica, Universit\'a dell'Insubria, Via Valleggio 11
Como, Italy }

\date{\today}

\begin{abstract}
We investigate the nature of the ground state of the one-dimensional 
$t-J$ model coupled to adiabatic phonons by use of the Lanczos technique at quarter filling.
Due to the interplay between electron-electron and electron-phonon 
interactions, the model undergoes instabilities toward the formation of lattice 
and charge modulations. Moderate on-site and intra-site electron-phonon couplings
lead to a competition of different spin-Peierls and dimerized states.
In the former case two electrons belong to the unit cell and we 
expect a paramagnetic band insulator state, while lattice dimerization leads to 
a Mott insulating state with quasi long range antiferromagnetic order.
The zero temperature phase diagram is obtained as a function of intra-site and inter-site 
electron-phonon couplings, analytically in the $J\to 0$ limit and numerically at finite $J/t$.
\end{abstract}

\pacs{75.10.Jm 05.70.Jk 75.40.Cx}
\maketitle
\section{Introduction}
The coupling between electrons and lattice degrees of freedom in 
strongly correlated systems has been the 
subject of several experimental and theoretical investigations showing that
low dimensional materials are prone 
to structural distortions driven by electron-phonon interaction\cite{bourjer,voit}. 
The most celebrated case is the Peierls instability, extensively studied
in weakly correlated models: the ground state of a one dimensional electron-phonon system 
is characterized by a lattice distortion which opens a gap at the
Fermi energy and gives rise to a charge modulation. 
In a half filled band, the density wave is commensurate to the lattice,
the unit cell doubles and the chain shows a spontaneous dimerized ground state.
When on site Coulomb repulsion cannot be neglected, the system displays 
competition between a Mott insulator state with strong antiferromagnetic correlations, 
favored by electron-electron interaction, and a paramagnetic band insulator state 
stabilized by lattice effects. 

Experimentally, there are indications for the importance of both
electron-electron and electron-phonon interactions in 
quasi-one dimensional systems. In particular there is a wide class 
of low dimensional materials, like the Bechgaard-Fabre salts series, where
the presence of these interactions leads to several modulated
phases such as: charge-density-wave (CDW), spin-Peierls (SP),
antiferromagnetic (AF), spin-density-wave (SDW), and even a 
high temperature superconducting state\cite{wilhelm}.

The physical mechanism which drives these different orderings is 
not fully understood yet. A satisfactory theory should be able to explain the 
origin and the stability of different patterns,
and the temperature scale at which they occur. In any case, 
the role of electron-phonon and electron-electron interactions 
will be instrumental for the  explanation of the presence of bond and charge order. 

By now, the physics of one dimensional (1D)
electron systems in a rigid lattice is well understood \cite{oned} 
in terms of the Luttinger Liquid (LL) paradigm:
a LL is a 1D paramagnetic metal characterized by 
gapless excitations both in the spin and in the charge channel.
In spin isotropic models, the low energy properties are determined 
by a single parameter denoted by $K_{\rho}$, in particular
$K_{\rho} < 1$ for repulsive interactions
and $K_{\rho} > 1$ in the attractive regime. The former case
is characterized by strong antiferromagnetic correlations, while 
in the latter dominant superconducting fluctuations are present.
Luttinger Liquid theory is 
an invaluable tool for the study of the effects of perturbations 
on an interacting 1D electron gas. Within this theoretical framework, it has
been shown that electron-lattice coupling, impurities and interchain 
interaction can destabilize the LL behavior opening a gap in the spin and/or 
charge excitation spectrum. A charge gap leads to an insulating state
whose properties depend whether the spin sector remains gapless or not.
A gapless spin spectrum leads to a 1D Mott insulator with (quasi) 
antiferromagnetic ordering, while a gap in the spin channel corresponds to 
a band insulator. In short range models, a spin gap with no charge gap  
preludes to a (quasi) superconducting state. 

Lattice effects in quasi one dimensional organic conductors
have been mostly investigated on the basis of the Hubbard model
with additional phonon couplings in adiabatic approximation
(Peierls-Hubbard model) \cite{mazclay,rierapoil}. 
Recent works extended the study to 
include quantum effects for the phonons \cite{seng}. However, the role of phonon
coupling in 1D models which may display attractive interactions,
like the $t-J$ model, received comparatively less attention \cite{rierapoil2,bispar}. 
The simultaneous presence of both bond and on-site (Holstein) phonons is likely to be an
essential ingredient for a faithful representation of quasi
1D organic materials. In this paper we explore systematically 
the effect of adding electron-phonon interactions to the $t-J$ model  
by including inter-site (bond) and on-site (Holstein) phonon 
couplings in the adiabatic approximation. We aim at the understanding of 
the qualitative features of the lattice periodicities which are stabilized and the sequence of 
modulations the model undergoes when the strength of phonon couplings is varied. 
\section{The model}
The $t-J$ hamiltonian provides one of the simplest models able to capture
the essential low energy physics 
of doped antiferromagnets. The electron dynamics is governed by a
competition between two terms: the kinetic contribution which describes
hopping between adjacent sites and the nearest neighbor Heisenberg 
interaction, which favors antiferromagnetic alignment of spins.
An infinite on site repulsion is also included in the model by means of 
a constraint. The phase diagram of the pure $t-J$ model 
has been numerically determined\cite{ogata} while 
analytic solutions are known only in the 
$J \rightarrow 0$ limit \cite{shiba} 
and at the supersymmetric point $J/t=2$ \cite{bares,bares1}.
At half filling the model reduces to the Heisenberg chain and 
when it is adiabatically coupled to the lattice via an arbitrarily weak perturbation,
the system undergoes spontaneous dimerization (spin-Peierls instability) 
leading to a lattice distortion with wave vector $q=2k_F$:
adjacent spins are paired in a singlet state and a spin gap in the excitation 
spectrum shows up. For electron density $n<1$ the pure $t-J$ model enters the Luttinger Liquid 
phase and, at larger $J/t$, phase separation sets in\cite{ogata}.

The one-dimensional $t-J$ model coupled with adiabatic phonons is defined by the
hamiltonian
\begin{eqnarray}
H &=& - \sum_{i} {\tonda{1-\delta_i}}
\tonda{\widetilde{c}_{i,\sigma}^{\dagger}\widetilde{c}_{i+1,\sigma}+ h.c.} + \nonumber \\
&+& J\sum_{i}\tonda{1-g\delta_i} \tonda{\gr{S}_i\cdot \gr{S}_{i+1}- \frac{1}{4}n_in_{i+1}} + \nonumber \\
&+& V\sum_{i} n_in_{i+1}+ \nonumber \\
&+& \sum_{i} n_iv_{i} + \frac{1}{2}K_B \sum_{i} \delta_i^{2} +\frac{1}{2}K_H \sum_{i} v_i^{2} 
\label{tjh}
\end{eqnarray}
where $\gr{S}_i$ are spin-$1/2$ operators at the site $i$,
$\widetilde{c}_{i,\sigma}^{\dagger}=c_{i,\sigma}^{\dagger}\tonda{1-n_{i,-\sigma}}$
are electron Gutzwiller-projected creation operators and
$n_i=\sum_\sigma c_{i,\sigma}^{\dagger}c_{i,\sigma}$ is the local electron density.
We have also set the bare hopping integral $t$ to unity thereby fixing the energy scale.
Hamiltonian (\ref{tjh}) includes both nearest neighbor superexchange coupling $J$ and 
Coulomb repulsion $V$.
The hamiltonian parametrically depends on the classical variables ${\delta_i}$ 
and ${v_i}$ which identify
the bond distortion and the amplitude of the internal site vibration, respectively.
The bond distortions are defined as 
$\delta_i= (u_i-u_{i-1})$, where $u_i$ is the displacement 
of the $\it{i}$ site from the equilibrium position, while the
on-site displacement $v_i$ corresponds to an effective parameter 
which combines several processes.
Clearly, the physical range where the hamiltonian (\ref{tjh}) may be used is
restricted to $|\delta_i| \ll 1$.
The first three terms correspond to the usual $t-J-V$ model coupled with adiabatic bond phonons,
the next one is the Holstein type interaction between the particles and the lattice, while 
the last two terms describe
the elastic deformation energy, $K_B$ and $K_H$ being the spring constants.

The $t-J$ model (\ref{tjh}) can be regarded as the strong coupling limit of the 
extended Hubbard model when $U\to\infty$. This leads to a definite relationship 
between the Hubbard and $t-J$ couplings: $J=4/(U-V)$ and $g=2$ \cite{rierapoil2}. 
However, it is known \cite{rice} that the $t-J$ model can be also considered as an {\sl effective}
one band hamiltonian unrelated to the Hubbard model. In this case, the 
superexchange coupling $J$ takes into account electronic processes involving other
degrees of freedom (like the Oxygen orbitals in Copper Oxide materials and in manganites
\cite{manganites}). The study of the effects of phonon coupling in the $t-J$ model allows to
analyze the role of magnetic interactions in stabilizing lattice distortions in
itinerant electron models. By varying the $J$ parameter we can explicitly tune the 
strength of antiferromagnetic coupling in the system: the comparison between the
(analytical) $J\to 0 $ limit and the (numerical) finite $J$ results will contribute 
to the understanding of the physical mechanisms leading to the observed periodicities. 
In the following we will therefore consider $J$, $V$ and $g$ as free parameters in the
$t-J$ hamiltonian (\ref{tjh}). 

The calculations have been performed at quarter filling $n=1/2$, 
which is an appropriate choice to mimic the behavior of the organic compounds. 
By use of Lanczos diagonalizations, we found the lattice distortions providing 
the lowest energy, without
any assumption on the periodicity of the broken symmetry ground state.
The extremum condition defining the optimal modulation can be formally expressed as
\begin{equation}
\frac{\partial {\langle H \rangle}}{\partial\delta_i}= 0; \qquad\qquad
\frac{\partial {\langle H \rangle}}{\partial v_i} = 0 
\label{self}
\end{equation}
The sum rule $\sum_i \delta_i=0$ reflecting the ring geometry of the model
has been implemented by use of a Lagrange multiplier, 
while no constraint on $\{ v_i\}$ has been imposed.
Here $\langle ...\rangle$ is the ground-state expectation value 
obtained by exact diagonalization using the Lanczos algorithm.
In order to find the bond length configuration which minimizes
the total energy $E(\graf{\delta_i}, \{v_i\})$, 
at fixed spring constants $(K_B, K_H)$, 
we iterate the set of equations (\ref{self}) until convergence is reached.

Following a previous investigation\cite{bispar} we chose open shell boundary 
conditions in order to mimic the gapless behavior of the (pure) $t-J$ model 
in the thermodynamic limit when studying a small size chain ($N\leq 16$).
\section{Analytical results}
In the special limit $J \to 0$ (and $V=0$) the problem can be solved analytically
even in the thermodynamic limit because the charge and spin degrees of freedom 
exactly factorize at all length scales \cite{shiba}. 
By setting $J=0$ in Eq. (\ref{tjh}) the hamiltonian reduces to the pure
kinetic contribution. In 1D the single occupancy constraint of the $t-J$ model
can be explicitly taken into account by promoting the spinful electron to
spinless fermions. The original electron problem at quarter filling then
maps into the adiabatic Su-Schrieffer-Heeger (SSH) model \cite{fra} for spinless
fermions at half filling with an additional 
local potential $v_i$ and elastic deformation energies. 

Being at half filling, 
the hamiltonian (\ref{tjh}) is unstable toward dimerization and
a gap in the energy spectrum opens for arbitrarily weak electron phonon coupling.
In the SSH adiabatic case ($v_i=0$) a dimerized lattice distortion pattern 
with uniform charge is stabilized and gives rise to a bond ordered wave (BOW).
In the pure Holstein case ($\delta_i=0$)
the lattice is undistorted and a CDW forms. A competition between BOW and CDW dimerizations 
is expected in the general case. By assuming a dimerization pattern of the form
$\delta_i=(-1)^i\delta$ and $v_i= (-1)^iv - v_0 $, the hamiltonian (\ref{tjh}) 
for $J\to 0$ becomes:
\begin{eqnarray}
H &=& - \sum_{i} {\tonda{1-\delta(-1)^i}}
\tonda{{c}_{i,\sigma}^{\dagger}{c}_{i+1,\sigma}+ h.c.} + \nonumber \\
&+& v\sum_{i}(-1)^i n_i +  \frac{N}{2}K_B  \delta^{2} + \nonumber \\
&+& \frac{N}{2}K_H  v^{2} - v_0 N_e +\frac{N}{2}K_Hv_0^{2}
\label{splessham}
\end{eqnarray}
where $N$ is the number of sites and $N_e=N/2$ is the number of particles.

\begin{figure}
\includegraphics[width=0.4\textwidth]{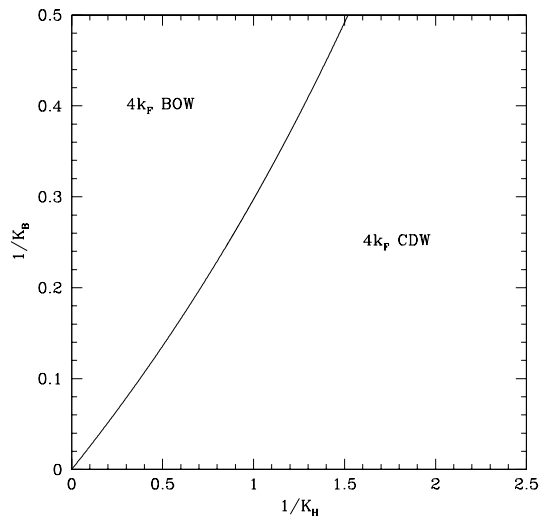}
\caption{\label{fig1} Zero temperature phase diagram of the $t-J$ model coupled to 
adiabatic phonons in the $J\to 0$ limit.}
\end{figure}

\begin{figure}
\includegraphics[width=0.4\textwidth]{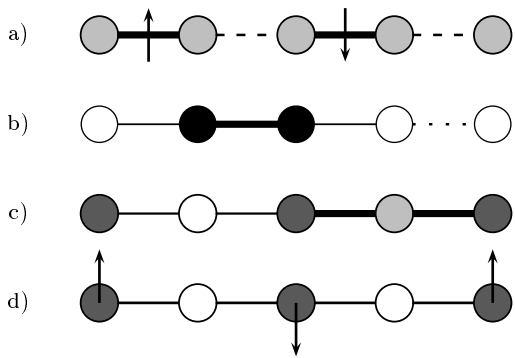}
\caption{\label{fig2}
Pictorial representations of the competing density waves in the ground state
of the $t-J$ model. 
a) $4k_F$ BOW two different bonds and uniform charge,
b) $2k_F$ BCDW tetramerized phase (I-S-I-W)
three different bonds and two different charges,
c) $4k_F$ CDW-SP (W-W-S-S) two different bonds and three different charges,
d) $4k_F$ CDW uniform bond and two different charges.
}
\end{figure}

In this case the hamiltonian can be easily diagonalized
by canonical transformations and the total energy, 
in the thermodynamic limit, has the form:
\begin{eqnarray}  
E = - N\int_{-\pi/2}^{\pi/2} &\phantom{+}& \frac{dq}{2\pi}
\quadr{4\tonda{\cos^2q+\delta^2\sin^2q}+v^2}^{\frac{1}{2}} + \nonumber \\
&+& \frac{N}{2}K_B  \delta^{2} + \frac{N}{2}K_H  v^{2} - \frac{N}{8K_H} 
\end{eqnarray}
The minima of the energy $E$ as a function of the 
dimerization parameters $\delta$ and $v$ always correspond to
pure BOW ($v=0$) or pure CDW ($\delta=0$) according to the values
of the elastic constants. The two modulations are respectively given  
by the equations:
\begin{eqnarray}
K_B &=& \int_{-\pi/2}^{\pi/2}\frac{dq}{2\pi} \: 
2\sin^2 q\tonda{\cos^2q+\delta^2\sin^2 q}^{-\frac{1}{2}}  \nonumber \\
K_H &=& \int_{-\pi/2}^{\pi/2}\frac{dq}{2\pi} \tonda{4\cos^2q+v^2 }^{-\frac{1}{2}}
\end{eqnarray}
Finally, the BOW/CDW phase boundary 
is obtained by equating the energies of the two distortions. The result is shown
in Fig.\ref{fig1}: the asymptotic form of the transition line for large elastic constants (i.e.
for weak distortions) is given analytically by $K_B=4K_H -2/\pi$.
We observe that at small $K_H$ 
a charge dimerization in the undistorted lattice ($4k_F$ CDW state: see Fig.\ref{fig2}d) 
prevails, while in the rest of the phase diagram a
dimerized lattice with uniform charge ($4k_F$ BOW state: see Fig.\ref{fig2}a) is stabilized.
In both cases a gap in the charge excitation spectrum is present
and the system is therefore an insulator. 

In order to characterize the electronic ground state of the model we observe that
even in the presence of adiabatic phonons the spin-charge factorization \cite{shiba} 
holds and therefore, following the analysis carried out for the strong coupling 
limit of the one dimensional Hubbard model \cite{parola}, the spin correlations can
be expressed as a convolution \cite{bispar}:
\begin{equation}
\langle \gr {S}_r\cdot \gr {S}_0 \rangle = 
\sum_{j=2}^{r+1}P^r_{SF}\tonda{j}S_H\tonda{j-1}
\label{spin}
\end{equation}
where $P^r_{SF}\tonda{j}$ is the probability of finding $j$ particles in $\tonda{0,r}$
with one particle in $0$ and another in $r$, evaluated in the ground state of the
dimerized spinless Fermi gas. If we set $N_r=\sum_{i=0}^r n_i$, the probability
$P^r_{SF}\tonda{j}$ can be formally expressed as
$P^r_{SF}\tonda{j} = \langle n_on_r\delta\tonda{N_r - j} \rangle$.

The asymptotic decay of Eq. (\ref{spin}),
in the presence of lattice dimerization and uniform charge 
(Fig.\ref{fig2}a) has been previously evaluated in Ref.\cite{bispar}:
\begin{equation}
\langle \gr {S}_r \cdot \gr {S}_0 \rangle \propto 
\frac{\cos(2k_Fr-\frac{\pi}{4})}{r}
\label{heisLATT}
\end{equation}
and shows antiferromagnetic quasi long range order. 
Here the Fermi wave vector is given by $k_F=n\pi/2 = \pi/4$.
An analogous scenario is expected in the CDW phase, where
the lattice is uniform but the charge is dimerized (see Fig.\ref{fig2}d).
In this case the spin-spin correlation function is not translationally invariant
and decays as:
\begin{equation}
\langle \gr {S}_r \cdot \gr {S}_0 \rangle \propto 
\frac{\cos(2k_Fr)}{r} 
\label{heisCAR}
\end{equation}
within each sublattice. This asymptotic result, valid for any charge dimerization,
can be obtained in a simple way in the limiting case where 
the electron density is zero in one sublattice (i.e. when $ v \to \infty$).
In this case $P^r_{SF}\tonda{j} =\delta_{j,r/2 + 1} $ which leads to (\ref{heisCAR})
via Eq. (\ref{spin}).
Therefore in both the BOW and CDW case, the $t-J$ model is characterized by
a charge gap together with a $1/r$ decay of spin correlations and 
diverging antiferromagnetic susceptibility, as in the 1D Heisenberg model. 
Such a scenario strongly suggests that, even for $J\to 0$, lattice coupling favors 
antiferromagnetism, which will be eventually stabilized when the inter-chain 
coupling, present in the three dimensional systems, is taken into account. 

Interestingly, this solution directly applies to the strong coupling limit of
the Hubbard model via the correspondence $J=4/U$. Therefore we expect that the 
phase diagram of Fig. \ref{fig1} faithfully represents the behavior of the 
$U\to\infty$ Hubbard model when only the on site Coulomb repulsion is considered 
(i.e. $V=0$). The presence of finite range interactions is likely to play an 
important role in stabilizing other periodicities. 

\section{Lanczos diagonalizations}
We now turn to the analysis of the full hamiltonian (\ref{tjh})
including the super-exchange term $J$ and both Holstein and Peierls lattice couplings.
Unfortunately for finite values of $J$ it is not possible to derive exact results in the
thermodynamic limit and then we investigate this model at quarter filling 
and vanishing magnetization. 
Results are given for a fixed value of magnetic and magnetoelastic couplings 
($J=1.4$, $V=0$ and $g=1$) 
in order to discuss the interplay between bond and charge distortions in the presence of 
antiferromagnetic
interactions. We note that, due to strong finite size effects, a large value of the
magnetic coupling $J$ is necessary to stabilize the possible symmetry broken phases.
Otherwise, lattice and charge modulations are present only for exceedingly small
elastic constants. However, we believe that the {\sl qualitative} features of the phase 
diagram at realistic values of $J$ are well reproduced by our small cluster calculations,
while we expect that the position of the phase boundaries and the amplitude of the lattice 
distortions will be severely renormalized in the thermodynamic limit. 

In the ground state we observe only two competing periodicities:
tetramerization ($q=2k_F$) and dimerization ($q=4k_F$).
In the former case two electrons per unit cell are present
and then the system will probably be
a paramagnetic band insulator with both charge and spin gap.
This regime corresponds to the spin-Peierls phase
experimentally found in the Fabre and Bechgaard salts series\cite{wilhelm}.
When a $q=4k_F$ modulation is stabilized, two sites belong to the unit cell 
thereby accommodating a single electron: the Umklapp scattering opens a charge gap
and the system behaves like a Mott insulator. 

\begin{figure}
\includegraphics[width=0.4\textwidth]{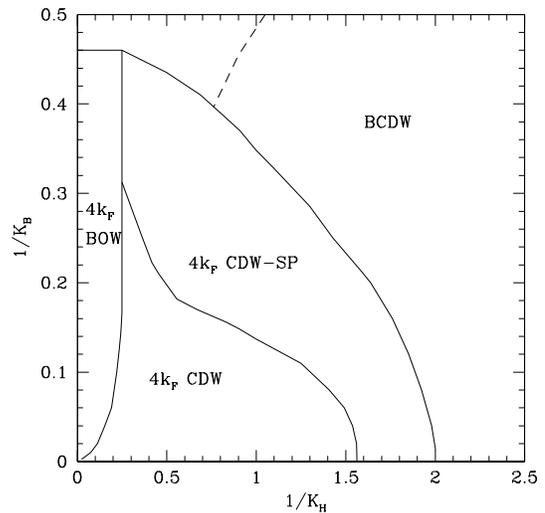}
\caption{\label{fig3}
Phase diagram of the t-J model coupled to adiabatic phonons (\ref{tjh})
obtained by numerical diagonalization of a $N \leq16$ site chain. The boundaries
of the dimerized bond order $4k_F$ BOW, the tetramerized $2k_F$ BCDW, 
the $4k_F$ CDW and uniform bond order, the $4k_F$ CDW-SP tetramerized bond-charge
order are schematically shown. The dashed line represents a crossover between two different 
{\sl metastable} states, as discussed in the text.}
\end{figure}

In Fig.\ref{fig3} we present a schematic ground state phase diagram in the $1/K_H$-$1/K_B$  
plane, showing the regions where the various broken symmetries dominate. 
Note that, as soon as both phonon couplings are significant,
mixed CDW-BOW ground-states are stabilized. 

All numerical results have been obtained by Lanczos diagonalization with no
assumption on the super-cell structure or on the periodicity of the ground state. This
limits the investigated cluster sizes to the range $N \leq 16$. In all cases the results 
do not depend on the starting bond configuration nor on the convergence requirement of
the adopted iterative method \cite{bispar}.
At large $K_H$, the numerical data show a narrow region 
where the ground state is a $4k_F$ BOW, as we found in the $J\to 0$ limit.
However, we suspect that in the thermodynamic
limit this phase disappears at finite $J$: our evidence for this comes
from few Lanczos diagonalizations carried out in a $N=20$ site chain 
\cite{sub} showing that the stability region of the $4k_F$ BOW phase
is considerably reduced, while the other phase boundaries hardly move.
Therefore we can conclude, as already suggested for the Peierls-extended Hubbard model\cite {mazclay},
that the $4k_F$ BOW ground state phase could be an artifact of the limited size chain.

Few calculations have been also performed for different parameter values. In all cases
we observed the same sequence of states shown in the phase diagram of Fig. \ref{fig3}
while only the phase boundaries are sensibly affected by the precise values of the couplings:
When $J$ is reduced, a uniform phase appears at large values of $K_H$ and 
the stability domains of the BOW phase gets larger. Remarkably, 
the qualitative appearance of the phase diagram remains unaltered
by tuning the magnetoelastic coupling $g$ and most of the changes can be
absorbed into a renormalization of the couplings, the effective $J$ being reduced when $g$
is increased. Similarly, by including the nearest neighbor Coulomb repulsion $V>0$
the rigidity of the system is enhanced and the effective elastic constants get larger.
However, no qualitatively different behaviors appear. 

For the investigated value of the super-exchange term $J$, 
all the various charge and bond modulations can be stabilized by
tuning the Holstein coupling at fixed $K_B$. 
Moving from weak to strong on-site coupling (i.e. by lowering $K_H$) three 
different types of structures are found: a $4k_F$ CDW, a $4k_F$ CDW-SP
and a $2k_F$ BCDW. In all these phases, 
the bond and charge density pattern can be simply parameterized as:
\begin{eqnarray}
\langle n_i \rangle &=& \frac{1}{2} + A_{4k_F} \cos{\tonda{\pi i}}+ 
A_{2k_F}\cos{\tonda{\frac{\pi}{2} i+\phi_c}}\nonumber\\
\delta_i &=& B_{4k_F}\cos{\tonda{\pi i}} +B_{2k_F}\cos{\tonda{\frac{\pi}{2}i + \phi_b}}
\label{dist}
\end{eqnarray}
At small on-site coupling ($K_H\to\infty$) $4k_F$ CDW instability prevails, 
while bonds remain uniform (see Fig.\ref{fig2}d). In this phase, the only
non vanishing amplitude in Eq. (\ref{dist}) is $A_{4k_F}$. 
Note that a previous investigation of the $t-J$ model without Holstein phonons 
\cite{bispar} showed no evidence in favor of the $4k_F$ CDW-SP and $4k_F$ CDW phases.
By lowering $K_H$, a tetramerized phase denoted $4k_F$ CDW-SP is obtained 
in the region of intermediate phonon couplings (see Fig.\ref{fig2}c): 
the charge ordering is characterized by a 
superposition of a $2k_F$ and a $4k_F$ density wave,
while the lattice modulation contains only a strong $2k_F$ component.
This regime corresponds to the  W-W-S-S bond sequence, 
where S and W respectively represent a weak (W) and strong (S) bond.
The charges order according to the sequence
1-0-1-0, where 1 (0) identifies a local densities higher (lower)
than the average ($n=1/2$). Note that, as shown in Fig.\ref{fig2}c,
the two low density (``$0$") sites inside the unit cell have different densities. 
The bond modulation is reproduced by setting $B_{2k_F}\neq 0$, $B_{4k_F}\simeq 0$
and $\phi_b=\pi /4$ in Eq. (\ref{dist}), while the charge pattern requires
$A_{2k_F}\neq 0$, $A_{4k_F}\neq 0$ and $\phi_c=0$. 
At small $K_H$ the ground state is characterized by 
a mixed $2k_F$ and $4k_F$ lattice order accompanied by a strong $2k_F$ charge modulation
(BCDW phase). The bond sequence is of the form I-S-I-W (where I stands for {\sl intermediate}, i.e. a 
moderately strong bond) and the charge order is the type of 0-1-1-0 (see Fig.\ref{fig2}b). 
The site centered BOW corresponds to 
both $B_{4k_F}\neq 0$ and $B_{2k_F}\neq 0$ with $\phi_b=0$, and the
$2k_F$ bond centered CDW is parameterized by $A_{2k_F}\neq 0$, $A_{4k_F}\simeq 0$ 
and $\phi_c=\pi/4$.  Notice that while the strong (S) and weak (W) bonds 
are affected by the strength of the on-site and bond couplings, the intermediate 
(I) bonds remain roughly constant throughout the BCDW region. 
By comparing these patterns to those obtained in the absence of Holstein phonons
\cite{bispar} we see that in the BCDW phase the S-W-S-W$'$ bond sequence is never stabilized
(except in a narrow region close to the phase boundary). However, by reducing the antiferromagnetic
coupling or by including a nearest neighbor Coulomb term $V$, a small BCDW region characterized
by the S-W-S-W$'$ sequence does appear, as already noticed in a previous investigation \cite{bispar}.
This suggests that in the small $J$ limit, where the $t-J$ model reduces to the strong coupling
regime of the Hubbard model, the S-W-S-W$'$ bond pattern shows up \cite{mazdix,rierapoil}.

The sequence of modulations displayed in the phase diagram of Fig.\ref{fig3}
does not change qualitatively with $K_B$ and
is maintained even at small bond phonon coupling, i.e. in the limit $K_B \to \infty$,
where the amplitude of the bond distortions becomes vanishingly small ($\delta_i \to 0$).
Contrary to the Hubbard model, 
the quantitative values of the site charge density wave 
$\langle n_i \rangle$ in the $4 k_F$ CDW-SP and $4 k_F$ CDW state 
are quite small, whereas in the BCDW state they are considerably larger.    
We could not stabilize, in our calculations, 
any $4 k_F$ CDW-SP or $4 k_F$ CDW states with large values of $\langle n_i \rangle$, 
nor any BCDW state with small $\langle n_i \rangle$.
\begin{figure}
\includegraphics[width=0.45\textwidth]{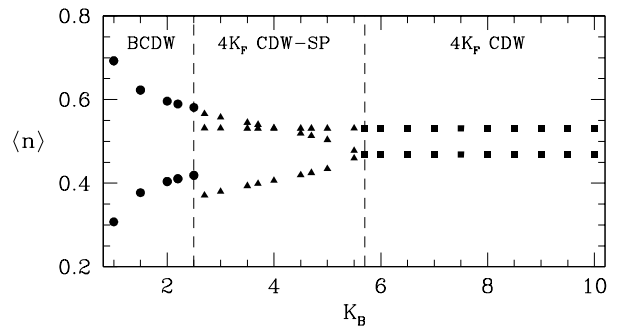}
\caption{\label{fig4}
Local densities in the unit cell as a function of $K_B$ 
at the given value $K_{H}=1$ }
\end{figure}
In Fig.\ref{fig4} we show a typical 
trend of the site charge density wave $\langle n_i \rangle$. The various points at each
$K_B$ provide the independent density values inside the unit cell of the superlattice
generated by the phonon distortions. 
Both the $4 k_F$ CDW-SP and the $4 k_F$ CDW regime are characterized by a 
substantially $K_B$-independent value of the highest density.
Instead, the lower density of the $4 k_F$ CDW state 
splits into two branches in the $4 k_F$ CDW-SP phase: the splitting increases continuously
even through the BCDW phase boundary. Similarly, by varying $K_H$ at fixed $K_B$ the 
size of the bond modulation and its periodicities are shown in Fig. \ref{fig4b} in 
the different regions of the phase diagram.
\begin{figure}
\includegraphics[width=0.45\textwidth]{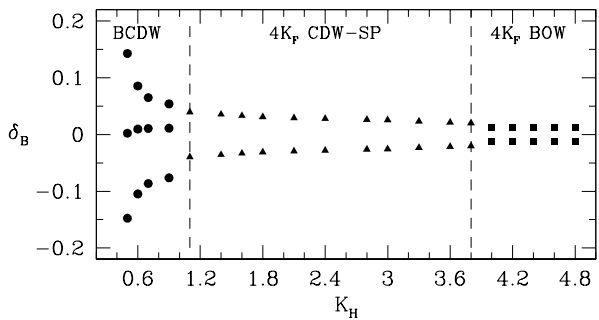}
\caption{\label{fig4b}
Bond strengths in the unit cell as a function of $K_H$ for $K_{B}=3$ }
\end{figure}

Whenever lattice tetramerization is stabilized, two electrons belong to the
unit cell and then the chain behaves as a paramagnetic band insulator, like
in the spin-Peierls phase. We believe that, because of the presence of
both charge and spin gap, this phase is stable toward 3D interchain coupling. 
While for lattice dimerization one electron belongs to 
the unit cell: electron repulsive interactions 
drive the system toward a  Mott insulator phase
and, if 3D coupling is allowed, antiferromagnetic order is favored. 

Our numerical study has been limited to investigate the ground state properties
of the hamiltonian (\ref{tjh}).
However a finite temperature analysis is relevant in order to apply
the results of our model to real materials. 
Some preliminary information about the temperature dependence can be inferred by
evaluating, via Lanczos diagonalizations, the 
energies of the lowest metastable state of the hamiltonian (\ref{tjh}) obtained by imposing 
a given periodicity of the bond/charge modulation. More precisely, let us
consider a point of the phase diagram  corresponding to a $2k_F$ modulation of the 
ground state, either of the BCDW or of the $4k_F$ CDW-SP type: We performed 
a self consistent Lanczos diagonalization by imposing the further constraint of 
$4k_F$ periodicity, i.e. of a two site unitary cell. In this way, the calculation
cannot converge to the actual ground state (which is characterized by a four site unitary cell)
but instead it selects a {\sl metastable state} identified as the lowest energy state 
with two site periodicity. 
\begin{figure}
\includegraphics[width=0.45\textwidth]{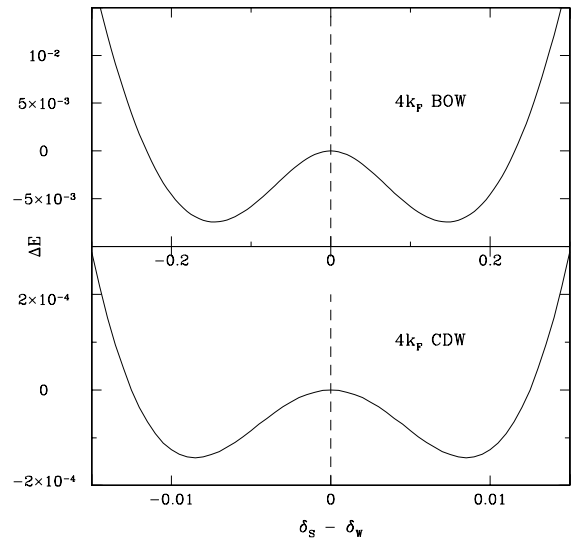}
\caption{\label{fig5}
Energy landscapes obtained by varying the bond length difference $\delta_S-\delta_W$
(see text) for $K_B=2$, $K_H=2$ (upper panel) and $K_B=5$, $K_H=1$ (lower panel).
}
\end{figure}
The nature of this metastable state changes throughout the phase diagram of Fig. \ref{fig3}: 
it is of the $4k_F$ CDW type (Fig. \ref{fig2}d) in the whole $4k_F$ CDW-SP region and
to the right of the dashed line in the BCDW phase, while it is of the $4k_F$ BOW type 
(Fig. \ref{fig2}a) in the remaining part of the BCDW region. The picture of the 
underlying energy landscape and the evaluation of the energy barrier separating the 
symmetry broken states can be also obtained by Lanczos diagonalizations. Starting from a
W-W-S-S configuration (Fig. \ref{fig2}c) we gradually changed the difference
$\delta_S-\delta_W$ between the two strong and the two weak lattice distortions in the unitary
cell, while letting the self consistent Lanczos algorithm optimize the vibrational parameters $v_i$.
A typical plot of the energy as a function of the bond asymmetry is shown in Fig. \ref{fig5} 
(lower panel). The broken symmetry $4k_F$ CDW-SP ground states correspond to the energy minima 
at non vanishing bond asymmetry, while the $4k_F$ CDW metastable state with uniform bonds is the
central maximum. When the thermal energy $k_BT$ is higher than the energy barrier $\Delta E$
we expect that thermal fluctuations will restore the full symmetry among the bonds in the
unitary cell stabilizing a $4k_F$ CDW state. Analogous calculation can be performed in the
other regimes. In Fig. \ref{fig5} (upper panel) the energy landscape corresponding to
the BCDW phase I-S-I-W is shown. Here, $\delta_S-\delta_W$ identifies the difference between the
strongest ($S$) and the weakest ($W$) bond, while the two other bonds (of equal strength) are kept
constant. The overall picture is quite similar to the previous case,
although the energy barrier is considerably higher. The metastable state at $\delta_S=\delta_W$
now corresponds to a dimerized (BOW) configuration with uniform charge: such a phase is then expected 
to represent the intermediate temperature regime in this region of the phase diagram. 
At higher temperatures a metallic uniform state is stabilized. 
The two scenarios discussed here are schematically shown in Fig. \ref{fig6}. Interestingly, 
the former case (panel a) corresponds to the ground state sequence
experimentally found in the Fabre salt series by increasing the temperature\cite{wilhelm}.
\begin{figure}
\includegraphics[width=0.47\textwidth]{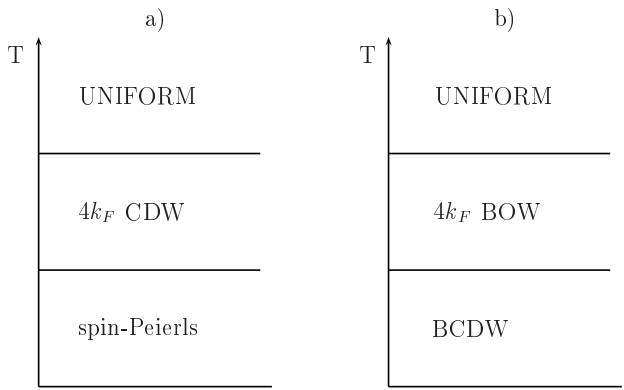}
\caption{\label{fig6}
Schematic expected sequences of phases obtained from a SP ground state 
(either BCDW or $4k_F$ CDW-SP) by increasing the temperature}
\end{figure}

\section{Conclusions}
We have investigated the $t-J$ hamiltonian
including both adiabatic Holstein and Peierls lattice couplings,
with amplitudes defined by the two independent parameters
$1/K_H$ and $1/K_B$, respectively. Exact analytical treatment of the $J\to 0$ limit
together with the implementation of a numerical method based on Lanczos technique
allowed to determine the ground state phase diagram 
as a function of the elastic constants at fixed value of the 
superexchange interaction $J=1.4$. For a weak on-site (Holstein) coupling (i.e. $K_H\to\infty$), 
two different dimerized regimes are stabilized: a $4k_F$ BOW and a $4k_F$ CDW
which display the typical features of a Mott insulator and suggest 
antiferromagnetic order in three dimensional samples.
In the former case the charge distribution remains uniform 
while the lattice is dimerized; in the latter case the
bond is uniform with dimerized charge. When $K_H$ gets smaller,
tetramerized phases appear displaying the typical behavior of the spin-Peierls regime: 
the $4k_F$ CDW-SP phase, for intermediate coupling, and the $2k_F$ BCDW for strong coupling.
The SP phase with 1-0-1-0 charge order has two kinds of bond and three distinct charges,
the bond distortion pattern is W-W-S-S and the distortion
makes the charges on the sites labeled ``0" unequal ($4k_F$ CDW-SP).
In contrast, the SP phase with 0-1-1-0 charge order is a BCDW phase,
with two different charges and three different bond order,
the bond distortion pattern is I-S-I-W. In the tetramerized phases
two electrons belong to the unit cell and then we expect a paramagnetic 
band insulator, and we believe that because of the presence of
both charge and spin gap, these two SP phases are stable toward 3D coupling 
and faithfully represent the low temperature phases of the model. 
The SP regimes give rise to two different scenarios at higher temperature:
by increasing the temperature, the system will explore metastable configurations 
corresponding to different bond/charge pattern before entering a paramagnetic 
undistorted phase. By examining the metastable states of our hamiltonian
we found two possibilities, one of them, the sequence showed in Fig.\ref{fig6}a, 
corresponds to the experimental phase diagram of Fabre salts in
the temperature-pressure plane \cite{bourjer,wilhelm}. Our study predicts that 
another sequence of phases, shown in Fig.\ref{fig6}b, should be possible
in quasi one dimensional materials characterized by larger $K_H$.

The effects of charge and bond ordering have been extensively investigated in the
extended Hubbard model\cite{mazdix,maztou} adiabatically coupled with the lattice. 
A ground state phase diagram obtained as a function of the electron-phonon
interaction shows the same periodicities that we found in our model, with the 
exception of the BCDW regime whose bond pattern follows the S-W-S-W$'$ sequence.
By lowering $J$ or by adding the nearest neighbor Coulomb repulsion $V$, the  S-W-S-W$'$
is recovered in a small window close to the BCDW phase boundary.  

This analytical and numerical study represents the first attempt to investigate the
interplay of bond and Holstein phonons in the one dimensional $t-J$ model. The
results show an overall similarity with respect to previous studies for the Hubbard 
model. However, even when the bond pattern is similar, the corresponding charge density 
wave displays different features between the two models due to the presence of a finite
range attraction, of magnetic origin, in the $t-J$ systems. A nearest neighbor attraction
smears the strong charge oscillations present, for instance, in the $4k_F$ CDW phase of the
Hubbard model \cite{mazclay}. A direct measure of charge and bond patterns in organic conductors 
will help in understanding the role of direct magnetic coupling in real materials.
Finally, another remarkable result which emerges form this work
is the counter-intuitive suppression of antiferromagnetic (quasi) long range order 
in strongly correlated models
when $J$ is {\sl increased}, due to the coupling to the lattice which drives a $2k_F$ 
instability, leading to a Spin-Peiers state.

We acknowledge fruitful discussions with V. Lante. Partial funding has been provided by MIUR 
through a PRIN grant.


\begin{thebibliography}{99}
\bibitem{sub} In order to limit the Hilbert space dimension, the allowed modulations have been
restricted to have period of four sites at most.
\end{thebibliography}
\end{document}